\documentclass[twocolumn,showpacs,prl]{revtex4}
\usepackage{graphicx}
\usepackage{bm}
\usepackage{psfig}
\usepackage{amsmath}
\usepackage{amssymb}

\begin{document}

\title{Observation of Fractional Quantum Effect at Even-denominator 1/2 and 1/4 Fillings in Asymmetric Wide Quantum Wells}
\author{J.~Shabani, T.~Gokmen, Y.T.~Chiu, and M.~Shayegan}
\affiliation{Department of Electrical Engineering, Princeton University, Princeton, NJ 08544, USA}
\date{\today}
\begin{abstract}
We report the observation of developing fractional quantum Hall states at Landau level filling factors $\nu = 1/2$ and 1/4 in electron systems confined to wide GaAs quantum wells with significantly $asymmetric$ charge distributions. The very large electric subband separation and the highly asymmetric charge distribution at which we observe these quantum Hall states, together with the fact that they disappear when the charge distribution is made symmetric, suggest that these are one-component states, possibly described by the Moore-Read Pfaffian wavefunction.
\end{abstract}


\maketitle

Even-denominator fractional quantum Hall states (FQHSs), observed in very high quality two-dimensional (2D) electron systems (ESs) have been enigmatic. In standard, single-layer 2D ESs confined to GaAs/AlGaAs heterojunctions or to narrow GaAs quantum wells, even-denominator FQHSs are observed only in the excited Landau levels, primarily at filling factor $\nu = 5/2$ \cite{Willett87,Pan99,Dean08}. It is yet not known whether the spin degree of freedom is necessary to stabilize this state \cite{Dean08}. If yes, then the 5/2 state could be described by a two-component (2C) Laughlin-Halperin ($\Psi_{331}$) wavefunction \cite{Halperin83}. But if it is stable in a fully spin-polarized 2D ES, then it is likely to be the one-component (1C) Moore-Read Pfaffian (MRP) state \cite{Moore91}. The latter is of particular interest as it is expected to obey non-Abelian statistics and have potential use in topological quantum computing \cite{Nayak08}.

The possibility of a 1C FQHS at even-denominator $\nu$ in the {\it lowest} Landau level, e.g. at $\nu = 1/2$, has been theoretically discussed in numerous publications \cite{Moore91,Greiter91,He93,Halperin94,Nomura04,Storni08,Papic09}. However, there has been no experimental evidence up to now that such a state exists. FQHSs at $\nu = 1/2$ have been seen in $bilayer$ ESs in either double \cite{Eisenstein92} or wide \cite{Suen92,Suen94,Manoharan96,Shayegan96} quantum well (DQW or WQW) systems; a FQHS at $\nu = 1/4$ was also observed recently in WQWs \cite{Luhman08,Shabani09}. In both DQW and WQW systems, when the interlayer tunneling is small, the 1/2 state is well described by the 2C $\Psi_{331}$ wavefunction; in this case the "components" are the layer indices or, alternatively, the two (symmetric and antisymmetric) electric subbands. In bilayer systems with strong tunneling (large symmetric-to-antisymmetric subband splitting, $\Delta_{SAS}$), on the other hand, the situation is unclear. According to theory a MRP FQHS, obeying non-Abelian statistics, can exist at $\nu = 1/2$ \cite{Greiter91,Halperin94,Nomura04,Storni08,Papic09}. Experiments, however, have shown that the $\nu = 1/2$ (and 1/4) FQHSs observed in WQWs are stable only when the overall charge distribution in the well is nearly symmetric ("balanced") and that the states disappear when the distribution is made asymmetric ("imbalanced") \cite{Suen94,Manoharan96,Shayegan96,Shabani09}. Moreover, for a given well width, the 1/2 and 1/4 FQHSs weaken and eventually disappear when the density is reduced and $\Delta_{SAS}$ is sufficiently increased. These observations were taken as evidence that these FQHSs are 2C \cite{Suen94,Halperin94,Manoharan96,Shayegan96,Shabani09}.

Here we report the observation of $\nu = 1/2$ and 1/4 FQHSs in WQWs with very significant charge distribution asymmetry and large subband separation. Ironically, when the charge distribution is made symmetric and the subband splitting is lowered, the states disappear. These observations suggest that these new FQHSs are 1C and are possibly of MRP origin.

Our structures were grown by molecular beam epitaxy and consist of GaAs WQWs bounded on each side by undoped AlGaAs spacer layers and Si $\delta$-doped layers. We present data on two samples, A and B, with well widths of 55 and 47 nm, respectively, and a mobility of $\simeq$ 250 m$^{2}$/Vs at a density of $n=2 \times 10^{11}$ cm$^{-2}$. An evaporated Ti/Au front-gate and a Ga back-gate were used to change $n$ and control the charge distribution symmetry. Magneto-resistance coefficients were measured in a van der Pauw geometry in $^{3}$He/$^{4}$He dilution refrigerators and in magnetic fields ($B$) up to 45 T.

It is instructive to first describe the ES confined to a WQW and how we change and monitor the charge distribution symmetry. When electrons at very low $n$ are confined to a modulation-doped WQW, they occupy the lowest electric subband and have a single-layer-like (but rather thick in the growth direction) charge distribution. As more electrons are added to the well, their electrostatic repulsion forces them to pile up near the well's walls and the charge distribution appears increasingly bilayer-like \cite{Suen94,Manoharan96,Shayegan96,Shabani09}. At such $n$ the electrons typically occupy the lowest two, symmetric and antisymmetric, electric subbands which are separated in energy by $\Delta_{SAS}$. An example of the charge distribution in such a system is given in Fig.~1(a) where we show the results of our self-consistent calculations for $n=1.72 \times 10^{11}$ cm$^{-2}$ electrons symmetrically distributed in a 55 nm WQW \cite{footnote1}. A remarkable property of the ES in a WQW is that both $\Delta_{SAS}$ and $d$ (the inter-layer separation), which characterize the inter-layer coupling, depend on $n$: increasing $n$ makes $d$ larger and $\Delta_{SAS}$ smaller so that the system can be tuned from a (thick) single-layer-like ES at low $n$ to a bilayer one by increasing $n$. This evolution with $n$ plays a decisive role in the properties of the correlated electron states in this system \cite{Suen94,Manoharan96,Shayegan96,Shabani09}. Equally important is the symmetry of the charge distribution in the WQW. For a fixed $n$, as the charge distribution is made asymmetric, the separation ($\Delta_{01}$) between the lowest two energy levels becomes larger than $\Delta_{SAS}$ and the system becomes increasingly single-layer-like. Figure~1(b) shows an example of the calculated charge distribution for the case where $n$ is the same as in Fig.~1(a) but electrons are transferred from one side of the WQW to the other side so that there is a layer density difference of $\Delta n = 7.8 \times 10^{10}$ cm$^{-2}$. As indicated in Figs.~1(a) and 1(b), the subband splitting increases from 25 K for $\Delta n=0$ to 41 K for $\Delta n = 7.8 \times 10^{10}$ cm$^{-2}$. Again, the symmetry of the charge distribution has a profound effect on the correlated states in a WQW \cite{Suen94,Manoharan96,Shayegan96,Shabani09}.

\begin{figure}[ht!]
\centering
\includegraphics[scale=0.6]{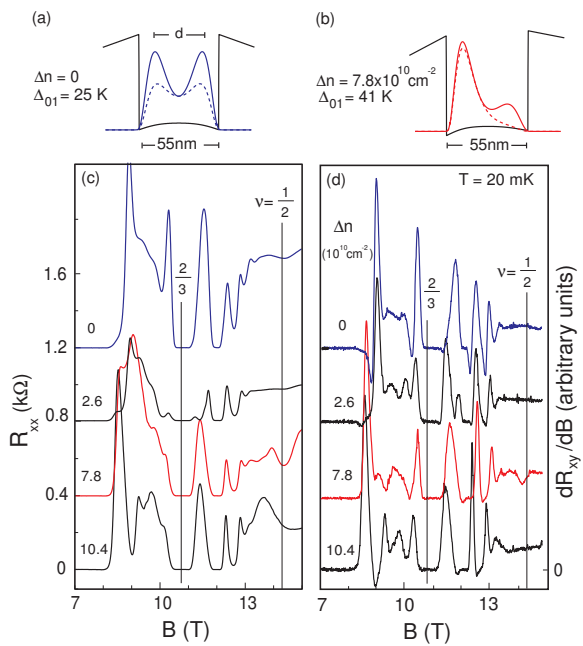}
\caption{(Color online) (a),(b) Self-consistently calculated (at $B = 0$) potential (black curves), total (blue/red solid curves), and the lowest subband charge distribution (blue/red dashed curves) for $n=1.72 \times 10^{11}$ cm$^{-2}$ electrons in a 55 nm-wide GaAs quantum well as the charge distribution is made asymmetric. (c),(d) Evolution of the magneto-resistance traces for sample A at a fixed $n=1.72 \times 10^{11}$ cm$^{-2}$ and for different $\Delta n$ as given on the left of each trace. We observe a developing FQHS at $\nu = 1/2$ for $\Delta n=7.8 \times 10^{10}$ cm$^{-2}$ (red traces). }
\end{figure}

Experimentally we control both $n$ and $\Delta n$ via applying voltage bias to front- and back-gates, and by measuring the occupied subband electron densities from the Fourier transforms of the low-field ($B \leq$ 0.4 T) magneto-resistance oscillations. These Fourier transforms exhibit two peaks whose frequencies are directly proportional to the subband densities. The difference between these frequencies is therefore a direct measure of $\Delta_{01}$. By carefully monitoring the evolution of these frequencies as a function of $n$ and, at a fixed $n$, as a function of the back- and front-gate biases, we can determine and tune the symmetry of the charge distribution \cite{Suen94,Manoharan96,Shayegan96,Shabani09}. Throughout this Letter we quote the experimentally determined values for $\Delta n$ and subband spacings ($\Delta_{01}$ or $\Delta_{SAS}$). The calculated subband spacings are in good agreement with the experimental values \cite{Suen94,Manoharan96,Shayegan96,Shabani09}.

\begin{figure}[ht!]
\centering
\includegraphics[scale=0.45]{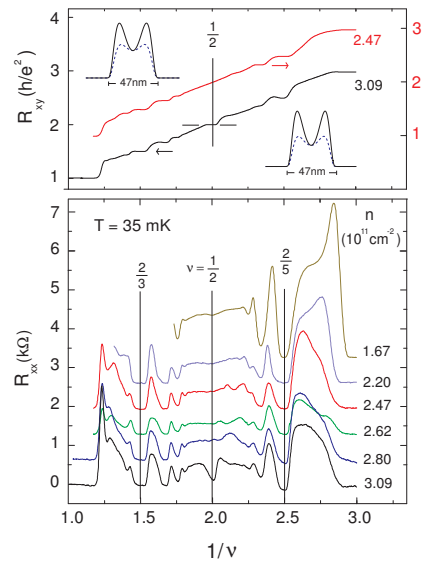}
\caption{(Color online) Evolution of magneto-resistance traces for sample B. The traces are shown for different total densities (indicated on the right) while the charge distribution is kept symmetric. As $n$ is decreased from 3.09 to $1.67 \times 10^{11}$ cm$^{-2}$, $\Delta_{SAS}$ increases from 34 K to 46 K. We observe a strong $\nu = 1/2$ FQHS only at the highest $n$. Calculated charge distributions for two representative densities are also shown. }
\end{figure}

\begin{figure*}[ht!]
\centering

\includegraphics[scale=0.75]{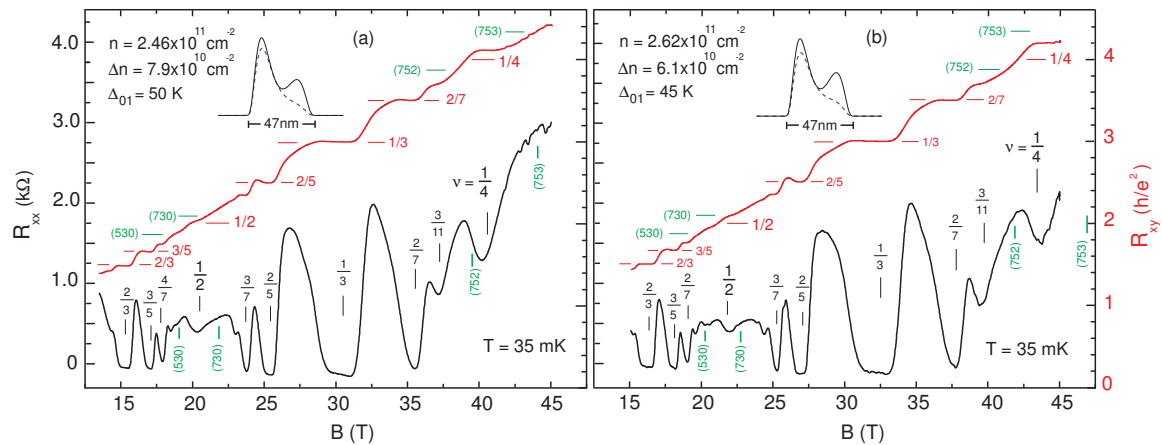}
\caption{(Color online) Magneto-resistance traces for sample B. The $expected$ positions of $R_{xx}$ minima for the commonly observed, odd-denominator FQHSs, as well as $\nu$ = 1/2 and 1/4, are marked. Also shown are the positions of $R_{xx}$ minima expected for the imbalanced 2C states (530), (730), (752), and (753). For $R_{xy}$ traces, the $expected$ positions of the plateaus for various FQHSs are indicated by horizontal lines. Calculated total and lowest subband charge distributions are also shown. }
\end{figure*}
In Fig. 1(c) we present the longitudinal ($R_{xx}$) and Hall ($R_{xy}$) resistance data vs $B$ for sample A; for the Hall data, we show the derivative $dR_{xy}/dB$ as it is more sensitive to the formation of a FQHS plateau. The traces are for different charge distributions at a constant $n = 1.72 \times 10^{11}$ cm$^{-2}$. It is clear that the data for the symmetric (or nearly symmetric) charge distribution do not show a FQHS at $\nu = 1/2$, but data taken at a significant charge imbalance ($\Delta n = 7.8 \times 10^{10}$ cm$^{-2}$) do. This is in sharp contrast to what has been seen in WQWs before \cite{Suen94,Manoharan96,Shayegan96,Shabani09}. We emphasize that the surprise here is that we observe a 1/2 FQHS as we make the charge distribution $imbalanced$ and $increase$ the subband separation.

We made measurements on sample B in a hybrid magnet with a maximum field of 45 T. The data are presented in Figs. 2 and 3. In Fig. 2 we keep the charge distribution balanced and change the total density by applying appropriate back- and front-gate biases. The $R_{xx}$ traces show the evolution of the state at $\nu = 1/2$, from an incompressible FQHS with a well developed $R_{xx}$ minimum and $R_{xy}$ plateau at high $n$ to a compressible state at low $n$. This evolution is consistent with previously observed trends, although the value of the parameter $\alpha = 0.13$ above which the 1/2 FQHS disappears is somewhat larger than what has been reported before for (wider) WQWs \cite{Suen94,Manoharan96,Shayegan96,Shabani09,Luhman08}; $\alpha = \Delta_{SAS} / (e^{2}/\epsilon l_{B})$ is defined as the ratio of the tunneling energy to the Coulomb energy. We would like to emphasize that sample B is the narrowest WQW in which a $\nu = 1/2$ FQHS has been observed up to now.

But the main surprises reveal themselves again when the ES is imbalanced. In Fig. 3(a) we show data for sample B at $n = 2.46 \times 10^{10}$ cm$^{-2}$ and $\Delta n = 7.9 \times 10^{10}$ cm$^{-2}$. Qualitatively similar to the data of Fig. 1, a FQHS at $\nu = 1/2$ emerges. Note that there is no such FQHS when the ES is balanced at this density (Fig. 2).

The very high magnetic fields provided by the hybrid magnet allow us to also explore lower filling factors in sample B. In Fig. 3(a), concomitant with the emergence of the $\nu = 1/2$ FQHS, we also observe a remarkably robust FQHS at another even-denominator filling, $\nu = 1/4$ \cite{footnote2}. In Fig. 3(b) we show data at a higher $n$; here, the front-gate bias was kept fixed and $n$ was raised by increasing the back-gate bias. Note that, compared to the data of Fig. 3(a), the FQHSs at $\nu = 1/2$ and 1/4 in Fig. 3(b) remain strong but their positions move up in magnetic field, consistent with the higher density.

The data presented in Figs. 1 and 3 provide evidence for the stability of $\nu = 1/2$ and 1/4 FQHSs in ESs with very large subband spacings and significantly asymmetric charge distributions. Moreover, at a fixed $n$, as the charge distribution is made symmetric and the subband spacing is lowered, the 1/2 and 1/4 states disappear, in contrast to previous observations. These characteristics strongly suggest that these are 1C states. We emphasize that at higher imbalances, the subband separation and charge distribution start to resemble those of single-layer 2D ESs in conventional GaAs/AlGaAs heterojunctions or (narrow) quantum wells which are known not to support a $\nu = 1/2$ or 1/4 FQHS. It is not surprising therefore that the states we observe disappear with larger magnitudes of charge imbalance.

Can these states have a 2C origin? The natural candidates would be the "imbalanced" $\Psi_{m_{b} m_{f} m}$ states where $m_{f} \neq m_{b}$ are odd integers and $m$ is an integer \cite{MacDonald90,Sawada98,footnote3}. In Table I we list the relevant characteristics of such states when their total filling is close to either 1/2 or 1/4. Near $\nu = 1/2$, the candidates are the (530) and (730) states, and near $\nu = 1/4$ the (752) and (753) states. Note that these states require unequal layer densities, with the front- to back-layer layer density ratio ($n_{f}/n_{b}$) ranging from 1.67 to 2.33. This is comparable to the experimental ratios where we observe the 1/2 and 1/4 states: $n_{f}/n_{b}$ for data of Figs. 1, 3(a) and 3(b) are 2.7, 1.9, and 1.6, respectively. The expected positions of $R_{xx}$ minima for the (530), (730), (752), and (753) states are marked with vertical lines in Figs. 3(a) and (b). It is clear that the observed minima are closer to the exact even-denominator 1/2 and 1/4 fillings than any of these 2C states.

\begin{table}

\caption{Characteristics of some of the 2C $\Psi_{m_{b} m_{f} m}$ states near $\nu$ = 1/2 and 1/4. }

\begin{tabular}{l l l c c c c c c c c}

\hline \hline
$m_{b}$ & $m_{f}$ & $m$ & & $\nu$ & & $\nu_{b}$ & & $\nu_{f}$ & & $n_{f}/n_{b}$ \\ \hline
5 & 3 & 0 & & 8/15  & & 1/5  & & 1/3  & & 5/3 \\ 
3 & 3 & 1 & & 1/2   & & 1/4  & & 1/4  & & 1 \\ 
7 & 3 & 0 & & 10/21 & & 1/7  & & 1/3  & & 7/3 \\ \hline 
7 & 5 & 2 & & 8/31  & & 3/31 & & 5/31 & & 5/3 \\ 
5 & 5 & 3 & & 1/4   & & 1/8  & & 1/8  & & 1 \\ 
7 & 5 & 3 & & 3/13  & & 1/13 & & 2/13 & & 2 \\ \hline \hline

\end{tabular}

\end{table}

In Fig. 3 we also mark the positions of the expected $R_{xy}$ plateaus. Near $\nu = 1/2$ the experimental traces at both densities show inflection points which are very close to the expected $2h/e^{2}$ value although the expected plateau for the (730) state is also quite close. Near $\nu = 1/4$ the value of the observed plateau in Fig. 3(a) is closer to $4h/e^{2}$ than to the values expected for the (752) and (753) states. The plateau in Fig. 3(b), however, is equally close to $4h/e^{2}$ and the (753) plateaus. It is worth noting that the {\it field positions} of the $\nu = 1/4$ plateaus in both Figs. 3(a) and 3(b), as revealed by plotting $dR_{xy}/dB$ vs $B$, agree very well with the observed positions of $R_{xx}$ minima. In other words, the observed plateaus are slightly $above$ the classical Hall resistance line. This deviation likely stems from the admixture of the $R_{xx}$ signal into $R_{xy}$, caused by a slight misalignment of the contacts in our van der Pauw shaped sample.

We believe that the imbalanced 2C states listed in Table I are unlikely to explain our data. First, the traces in Figs. 1 and 3 exhibit $R_{xx}$ (and $dR_{xy}/dB$) minima at several odd-denominator fillings whose field positions are consistent with what is expected in a simple, {\it 1C} 2D ES. Second, there is no obvious reason why the particular imbalanced 2C states listed in Table I should be stable. This is especially true for the states near $\nu = 1/4$ where the 2C candidate states have very low layer fillings. For example, the (753) state has layer fillings 2/13 and 1/13 and the lower density layer has intra-layer correlations similar to a single-layer 2D ES at $\nu = 1/7$. The stability of such a state is highly unlikely, given the extremely weak nature of the $\nu = 1/7$ FQHS in the highest quality, single-layer samples. Third, and most importantly, suppose, e.g., that the (753) is the FQHS we observe near $\nu = 1/4$. As we bring the ES to balance (at a fixed $n$), we lower the subband spacing. This should favor the 2C states and we would therefore expect to observe a 2C (553) FQHS at $\nu = 1/4$ when the ES is balanced. This is contrary to our observations \cite{footnote2}. The same argument could be made for the $\nu = 1/2$ FQHS.

As discussed earlier, the possibility of 1C, even-denominator FQHSs in single-layer or bilayer electron systems has been theoretically proposed \cite{Moore91,Greiter91,Halperin94,Nomura04,Storni08,Papic09}. In particular, in numerical studies, a significant overlap of the calculated ground state with the MRP wavefunction is taken as evidence for the viability of the MRP state. The experimental results presented here demonstrate the stability of 1/2 and 1/4 FQHSs in strongly asymmetric WQWs in a relatively narrow range of parameters such as well-width, electron density, and charge imbalance. More extensive measurements are likely to unravel the parameter range where these states are stable more comprehensively. But our results already challenge the theories by providing a set of parameters for which the stability of the $\nu = 1/2$ and 1/4 MRP states, or possibly other, yet unknown FQHSs, could be tested.

We thank the NSF (DMR-0904117 and MRSEC DMR-0819860) and DOE for support, J-H. Park, T. Murphy, G. Jones and E. Palm for experimental help, and B.A. Bernevig, F.D.M. Haldane, B.I. Halperin, L.N. Pfeiffer, D. Sheng, and E. Tutuc, for illuminating discussions. Part of our work was performed at the National High Magnetic Field Laboratory, which is supported by the NSF (DMR-0654118), the State of Florida, and the DOE.


\begin{thebibliography}{99}
\bibitem{footnote1}
All our self-consistent calculations were performed at $B = 0$ where the electrons occupy the lowest two subbands. At $\nu < 1$, in a single-particle picture the electrons should all reside in the lowest subband. In an interacting system, however, the Coulomb energy can far exceed the subband splitting and allow the antisymmetric subband to mix into the correlated ground state to lower its energy and a 2C state could ensue. Therefore, throughout this Letter we show our calculation results for the lowest subband and the total charge distributions.
\bibitem{footnote2}
The traces taken at balance at $n = 2.46$ and $2.62 \times 10^{11}$ cm$^{-2}$ for sample B do not show $\nu = 1/4$ FQHSs.
\end{thebibliography}
\end{document}